# Research Citations Building Trust in Wikipedia

Mike Taylor (University of Wolverhampton, Digital Science) 0000-0002-8534-5985 (Corresponding Author)
Carlos Areia (University of Coventry, Digital Science) 0000-0002-4668-7069
Kath Burton (Routledge, Taylor & Francis) 0000-0001-7785-9604
Charles Watkinson (University of Michigan Press) 0000-0002-9453-6695

## Abstract

The use of Wikipedia citations in scholarly research has been the topic of much inquiry over the past decade. A cross-publisher study (Taylor & Francis and University of Michigan Press) convened by Digital Science was established in late 2022 to explore author sentiment towards Wikipedia as a trusted source of information. A short survey was designed to poll published authors about views and uses of Wikipedia and explore how the increased addition of research citations in Wikipedia might help combat misinformation in the context of increasing public engagement with and access to validated research sources.

With 21,854 surveys sent, targeting 40,402 papers mentioned in Wikipedia, a total of 750 complete surveys from 60 countries were included in this analysis. In general, responses revealed a positive sentiment towards research citation in Wikipedia and the researcher engagement practices. However, our sub analysis revealed statistically significant differences when comparison articles vs books and across disciplines, but not open vs closed access. This study will open the door to further research and deepen our understanding of authors perceived trustworthiness of the representation of their research in Wikipedia.

## Introduction

The use of Wikipedia citations in scholarly research has been the topic of much inquiry over the past ten years. Studies determining the extent to which Wikipedia citations appear in books and journal articles have questioned the appropriateness of using Wikipedia as a source [1] in academic research while others have noted the value of engaging with Wikipedia more generally in combatting the spread of misinformation [2].

The growing body of literature about "academic Wikipedia" [3] suggests that the relevance of Wikipedia citations in relation to research impact is growing. Furthermore, publications exploring the accuracy of Wikipedia citations within scientific publications continue to increase (see Figure 1 below, source Altmetric); illustrating the important role of Wikipedia as an information literacy tool [4].

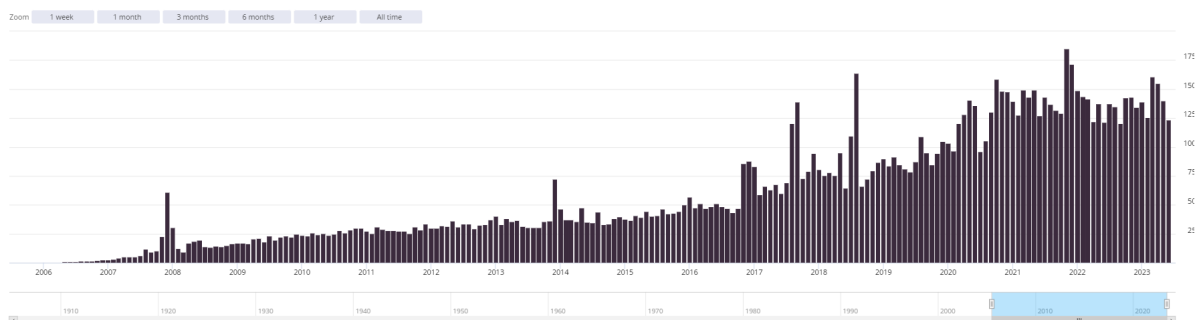

Figure 1: Source: Altmetric accessed 20 June 2023, WikipOI publication years 2014-2023

*Context of the 'academic Wikipedia' community and combating spread of misinformation.*

As the Wikimedia Foundation also notes, there is a "thriving movement" [5] in community sentiment towards improving the accuracy and trustworthiness of articles in Wikipedia. The community sentiment towards improving the accuracy of Wikipedia is reflected by the number of academic communities running edit-a-thons and dedicated conferences that are now emerging to connect scholars, professionals and the wider Wikipedia community [6] to develop partnerships for increasing how knowledge is represented and shared via Wikipedia.

Movements within the library and information studies community that support engagement with, and interrogation of Wikipedia are also increasing. Writing in *Leveraging Wikipedia: Connecting Communities of Knowledge* [7] Merillee Proffitt highlights the role of the "Wikipedian" - volunteer editors and supporters of Wikipedia - as central to ensuring that information held in Wikipedia remains credible. A role that has extended to residencies within libraries [8] and at publishers [9], indicating the significance of editing not only to improve the accuracy of source citations in publicly available information, but as important research and learning tool.

While initiatives like Wikithons and Residencies are useful engagement activities, the EZProxy Access now provided to a select group of Wikipedia Editors via The Wikipedia Library (and adopted by a growing number of scholarly publishers) is increasing the number of research citations to validated publications in Wikipedia. While the provision of access to vetted Wikipedia Editors via The Wikipedia Library (TWL) is instrumental in increasing the addition of accurate research citations, the effect of Open Access (OA) publication on citation accuracy in Wikipedia is an increasingly significant factor in establishing trust. Yang et al, 2023 in *Wikipedia and Open Access* [10] found that OA publication is likely to increase citation in Wikipedia, specifically for articles that have been recently published and yet to accrue many citations.

A 2020 published study by Nicholson et al [11] posed the question in relation to the >6.6m articles now published in Wikipedia (as of June 2023): "just how reliable are the sources cited by Wikipedia articles, particularly with respect to scientific topics?" The study found that research citations in Wikipedia "are more than twice as likely to be supported than the scientific literature in general" and that "the citation context offers a more complete picture, potentially affecting decisions by everyday readers and choices of editors"[11]. The findings from these two studies offer a powerful endorsement of Wikipedia's role in increasing the accuracy of citation to validated research in a publicly available source of information. Yet, there are no studies assessing the published authors' thoughts on Wikipedia citations to their articles or books.

Wikipedia plays a major role as an audience interface. As such ensuring Wikipedia's accuracy is of great importance and citation verifiability is key to Wikipedia's credibility as a trusted source information, but guidelines for the use of Wikipedia citations are outdated. Wikipedia demands reliable, published sources, e.g. peer-reviewed books and journal articles. Through initiatives like The Wikipedia Library (TWL), academic publishers are opening up their content to trusted TWL editors to add verifiable citations to Wikipedia records.

During 2023, a group of data scientists and scholarly publishers drawn from Digital Science, Routledge, Taylor & Francis and University of Michigan Press conducted an observational study into researchers' views on the accuracy of Wikipedia citations to their published works. This study set out to gather author sentiment towards Wikipedia as a trusted source of information. And explore how citation in Wikipedia might help combat misinformation in the context of increasing public engagement with and access to validated research. What value this presents to the researcher are key considerations of the study.

## Study Objectives

To help explore this, the study set out the following primary and secondary objectives:

**Primary:** To understand researchers' views on the accuracy and trustworthiness of Wikipedia in representing the outcomes of their research to a broader audience
**Secondary:**
1. Sub-group analysis of the primary outcomes by:
    a. Use of Wikipedia type (eg. research, teaching, etc…)
    b. Discipline
    c. OA status
2. Narratively explore and report free text answers.
3. Obtain a pool of researchers who we can follow up with to explore the topic of trust in Wikipedia and public engagement in more detail.

# Methods

## Design

This is a prospective observational study, with a cross-sectional data review of Wikipedia content and prospective contact of selected wiki pages cited researchers, referred to as "participants" from this point forward in the manuscript.

The survey was constructed in both Alchemer (T&F dataset) and Qualtrix (UMP) and a sample set of responses produced in each system prior to launch in order to set up a script to run the code for future analysis. The survey participants were contacted directly with an invitation to complete a survey via link embedded in the email. There was no open platform for anyone to participate in the survey and there was no mandatory requirement to complete the survey. No incentives were offered.

### Quality control and Ethical Statement

[https://www.equator-network.org/reporting-guidelines/improving-the-quality-of-web-surveys-the-checklist-for-reporting-results-of-internet-e-surveys-cherries/](https://www.equator-network.org/reporting-guidelines/improving-the-quality-of-web-surveys-the-checklist-for-reporting-results-of-internet-e-surveys-cherries/)
The study does not require IRB approval.

## Data collection

Using Dimensions and Altmetric (Digital Science) data, an initial dataset was built to identify Wikipedia mentions from a pre-defined list of publications. This list was created using the following rules:
- Published by Taylor and Francis (T&F) or University of Michigan Press (UMP)
- Published in the last 10 years

One Wikipedia mention per publication prioritised: by the most recent English language Wikipedia mention. If there were no English language Wikipedia mentions then the most recent citation was used.

According to Altmetric data a total of 3,966,439 published outputs are cited by Wikipedia. The study prioritised citations in English Wikipedia and targeted 40,402 of the published items cited by Wikipedia articles. The survey therefore targeted about 1% of the Wikipedia cited content available in Altmetric.

This dataset contained the following categories:

**Publication:**
1. Publication ID
2. Publication title
3. Publication year

**Wikipedia related:**
4. Wiki language
5. Wiki URL

**Researchers:**
6. Researcher ID
7. First and last name
8. ORCID
9. First publication year

**Disciplines:**
10. FoR-based 6 disciplines [Social Sciences, Humanities, Medical and Health Sciences, Life Sciences, Physics and Mathematical Sciences, Engineering and Technology). Note that papers can have more than one discipline.

## Participant contact

Using the dataset, the Altmetric team provided both T&F and UMP with respective DOIs of the cited publications. Both T&F and UMP proceeded to extract the respective list of authors for each publication and to invite them to respond to our short survey. For each publication, we picked a single citing Wikipedia page (prioritised by the most recent English language Wikipedia mention, and if there were no English Wikipedia mentions, the most recent one) for the authors to examine and comment on.

Each survey participant was informed that the data collected during the survey would be anonymised and used for the purposes of analysis. Participants were invited to share their contact details if they wished to be contacted after the survey had completed. No personal data was extracted from the survey participants' responses and all survey responses were stored in a secure part of the project leads' servers only accessible to the project leads.

## Survey

The survey consisted of seven questions, one multiple choice, four being a Likert scale (scoring from 'strongly disagree', 0, to 'strongly agree', 10), one binary (Yes/No), and one question in free-text (Table 1). The data was collected between December 2022 and January 2023. The questions were distributed over no more than two pages systems used to circulate the survey.

| N | Question | Type |
|---|---|---|
| 1 | For what purposes do you use Wikipedia in your own research? Choose all options that apply. | Multiple choice |
| 2 | The citation accurately represents the content of my publication. | Likert (0-10) |
| 3 | This Wikipedia page is a reasonable representation of its subject. | Likert (0-10) |
| 4 | I would recommend this Wikipedia page to a colleague in the field. | Likert (0-10) |
| 5 | I would recommend this Wikipedia page to someone not in the field who wishes to understand this subject. | Likert (0-10) |
| 6 | Would you be willing to be contacted again and to be part of a bigger | Binary (Yes/No) |

| | project on Wikipedia trust and public engagement? | |
|---|---|---|
| **7** | Anything to add about your response? | Free text |

Table 1: Survey Questions

# Free text analysis

The survey design incorporated an optional free-text question (Q7). Placed at the end of the survey, this question was designed to solicit additional comments from respondents about the choices provided in the previous questions in the survey.

Free text data analysis consisted in a two-step process to evaluating comments was devised:

1. High-level grouping of comments based on similarity (e.g. "The main reason I would not recommend this is because it is too short.")
2. Descriptive coding and definitions created based on high-level grouping (e.g. Inaccurate Information: wrong image, wrong data, incorrect citation, typos)

This process produced a list of 10 descriptive codes that were then applied to the qualitative data received. These are listed in the table 2 below.

| **Code** | **Definition [example]** |
|---|---|
| Irrelevant to our study | thank you, no, I'm too busy |
| Unable to comment fully | unfamiliar language, lack of subject expertise |
| Inaccurate information | wrong image, wrong data, incorrect citation, typos |
| Inaccurate emphasis | lack of objectivity, not well written, impact of study mis-represented, could do better, could include a different citation as research has moved on |
| How to use Wikipedia | I don't understand…; how do I edit; should I include citations |
| Editing the Wikipedia entry | specific ask to please correct citation, spelling, typos |
| Personal engagement with Wikipedia | I support them, I am an editor, I tell my students... |
| Benefits of Wikipedia | Summary, connect to public, |
| Disbenefits of Wikipedia | perpetuates inaccuracies, poor citation practice |
| Serious/legal implications | slander accusation, trigger warnings missing |

Table 2: Free-text coding and definitions applied to Question 7

The free-text Q7 was an optional part of the survey and did not require any data from respondents for them to complete the survey. Some respondents chose to add some simple comments (e.g. no, thank you). These comments were coded as "Irrelevant to our study" to simplify the qualitative analysis and accounted for around 2% of all free text responses provided.

# Data and Statistical Analysis

## Data Analysis

The initial dataset to identify relevant manuscripts and Wikipedia pages was built using Altmetric and Dimensions data inside Google BigQuery (GBQ). Surveys were built and shared to participants using Alchemer (for T&F) and Qualtrics (for UMP). Final survey datasets were then aggregated using R [12]and the tydiverse library [13]
Data was explored both in R, GBQ and Datastudio.

## Statistical Analysis

Descriptive statistics will be presented in raw counts, proportions, means and respective standard deviations, where appropriate. Inferential statistical analyses were conducted to test hypotheses, t-tests were employed to compare the means between two groups, while Analysis of Variance (ANOVA) was used to compare means among three or more groups.

# Results

A total of 21,854 surveys were sent, 1154 started (5.3% response rate) and 750 completed (65.0% completion rate), this included 740 participants from 60 different countries, only completed surveys were included in the analysis. A total of 727 unique publications had at least one author fully responding to the survey, a breakdown on the publication disciplines can be found in Table 3:

| Discipline | Number of | Number of surveys received |
|---|---|---|

|                                   | publications |     |
|-----------------------------------|--------------|-----|
| Social sciences                   | 231          | 232 |
| Humanities                        | 221          | 225 |
| Life Sciences                     | 159          | 168 |
| Physical and Mathematical Sciences | 99          | 99  |
| Medical and Health Sciences       | 87           | 90  |
| Engineering and Technology        | 36           | 35  |

Table 3: Breakdown of respondents by Fields of Research (FoR) aggregated discipline. To note is that one publication may be labelled as more than one discipline.

Most of the answers to Question 1 ("For what purposes do you use Wikipedia in your own research?") use Wikipedia for teaching (344; 46%) and research (208; 28%), this was consistent across disciplines (Figure 2):

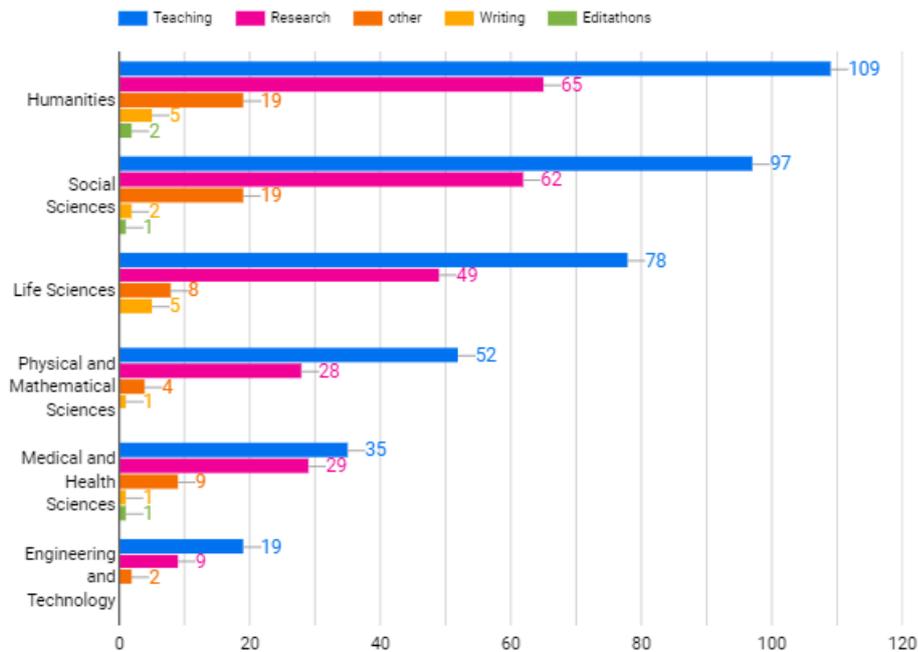

Figure 2: Uses of Wikipedia by FoR subject

The results of the survey indicate that the participants generally agreed with the statements in Question 2," The citation accurately represents the content of my publication." (mean = 7.65, SD = 2.56), Question 3, "This Wikipedia page is a reasonable representation of its subject." (mean = 7.03, SD = 2.78), and Question 4, "I would recommend this Wikipedia page to a colleague in the field." (mean = 7.43, SD = 2.27). However, Question 5, "I would recommend this Wikipedia page to someone not in the field who wishes to understand this subject.", was broadly supported, it received a slightly reduced agreement level (mean = 6.35, SD = 2.91).

## Sub-analysis

### Responder use

To compare responder type we have only selected individuals that selected either Teaching or Research (Table 4). All other options had too small sample size for any comparison. There is a statistically significant difference on citation representation (Q3) between Teachers and Researchers, with higher Trust from Teachers ($p = 0.018$). The results by responder type are noted in the table below.

| Group | N | Q2: Accuracy | Q3: Representation | Q4: colleague recommendation | Q5: public recommendation |
|---|---|---|---|---|---|
| **Responder Type** | | | | | |
| *Teaching* | 226 | 8.00±2.52 | 7.74 ±2.37[a] | 7.11 ±2.73 | 7.53 ±2.69 |
| *Research* | 136 | 7.70±2.27 | 7.36±2.18[a] | 6.64±2.60 | 7.16±2.67 |
| **Papers vs books** | | | | | |
| *Article* | 640 | 7.70±2.54 | 7.53±2.23[b] | 6.57±2.84[c*] | 7.24±2.68[d*] |
| *Book* | 100 | 7.34±2.69 | 6.88±2.48[b] | 5.32±3.10[c*] | 6.01±3.07[d*] |
| **OA Status** | | | | | |
| Closed | 474 | 7.59±2.57 | 7.45±2.35 | 6.47±2.94 | 7.15±2.82 |
| Open [all]: | 266 | 7.76±2.56 | 7.40±2.15 | 6.17±2.85 | 6.85±2.69 |
| Green | 112 | 7.84±2.66 | 7.27±2.36 | 6.15±2.91 | 6.79±2.84 |
| Hybrid | 70 | 7.87±2.35 | 7.33±2.11 | 5.88±3.13 | 6.64±2.92 |
| Gold | 55 | 7.58±2.51 | 7.77±1.81 | 6.77±2.47 | 7.31±2.14 |
| Bronze | 23 | 8.08±2.31 | 7.23±1.77 | 6.12±2.69 | 6.73±2.24 |
| Unlabelled | 6 | NA | NA | NA | NA |
| **Disciplines** | | | | | |
| SS | 232 | 7.74±2.37 | 7.38±2.13 | 5.85±2.94[e,f*] | 6.70±2.73[j] |
| H | 225 | 7.39±2.77 | 7.13±2.37 | 5.95±2.90[g,h*,i] | 6.72±2.90[k] |
| LS | 168 | 7.86±2.89 | 7.68±2.24 | 6.84±2.89[e, g] | 7.39±2.80 |
| PMS | 99 | 7.91±2.28 | 7.86±1.99 | 7.32±2.49[f*,h*] | 7.91±2.23[j,k,l] |
| MHS | 90 | 7.48±2.90 | 7.44±2.72 | 6.91±2.93[i] | 7.32±2.84[l] |
| ET | 35 | 8.00±2.54 | 7.86±1.72 | 6.81±2.67 | 7.41±2.01 |

Table 4: Segmented results showing statistically significant difference in comparison between Research and Teaching. [a-l] statistically significant between each other ($p < 0.05$), [a-l*] highly statistically significant between each other ($p < 0.001$). ET: Engineering and Technology, H: Humanities, LS: Life Sciences, MHS: Medical and Health Sciences, NA: Not Applicable, PMS: Physical and Mathematical Sciences, SS: Social Sciences

### Papers versus books

These findings suggest that the type of publication may influence the recommendation scores, with the accuracy of citations to articles receiving higher public and colleague recommendation scores than those to books ($p < 0.001$). Subject representation was also statistically significant between articles and books, also favouring articles ($p = 0.002$). Citation accuracy was not different (Table 4).

### OA status

Our analysis does not show a statistically significant difference between Open and Closed publications. When analysing the different OA types, gold OA showed higher scores for most questions, however, none of these are statistically significant (table 4). The OA status of publications referenced in the survey is included below for illustration purposes only.

## Disciplines

The dataset includes responses from six disciplines: Engineering and Technology, Humanities, Life Sciences, Medical and Health Sciences, Physical and Mathematical Sciences, and Social Sciences. Each discipline has a different number of responses, ranging from 35 for Engineering and Technology to 233 for Social Sciences (Table 4).

The mean public recommendation scores vary across disciplines, with Physical and Mathematical Sciences having the highest mean score (approximately 7.93) and Humanities having the lowest (approximately 6.7). Similar trends are observed for the mean scores of colleague recommendation, subject representation, and citation accuracy.

An ANOVA test was conducted to compare the mean public recommendation scores across the disciplines. The results indicate a statistically significant difference in the mean scores ($p < 0.05$). A Tukey HSD post-hoc test was then conducted to determine which specific disciplines have significantly different mean scores. The post-hoc test results highlight many significant differences. For example, it indicates that the Physical and Mathematical Sciences, when compared to Humanities and Social Sciences, have significantly higher public ($p < 0.05$) and colleague ($p < 0.001$) recommendation.

## Free-text results

Approximately 24% of the participants' response were to the free-text question (Q7). Several comments received (5.2% of all responses) were irrelevant to our study (e.g. Thanks!) and have not been discussed any further. The largest proportion of viable categorised responses (4.4%) relate to the lack of source publication context ("inaccurate emphasis") in the Wikipedia article. We discuss citation *accuracy* vs citation *emphasis* in the discussion section below.

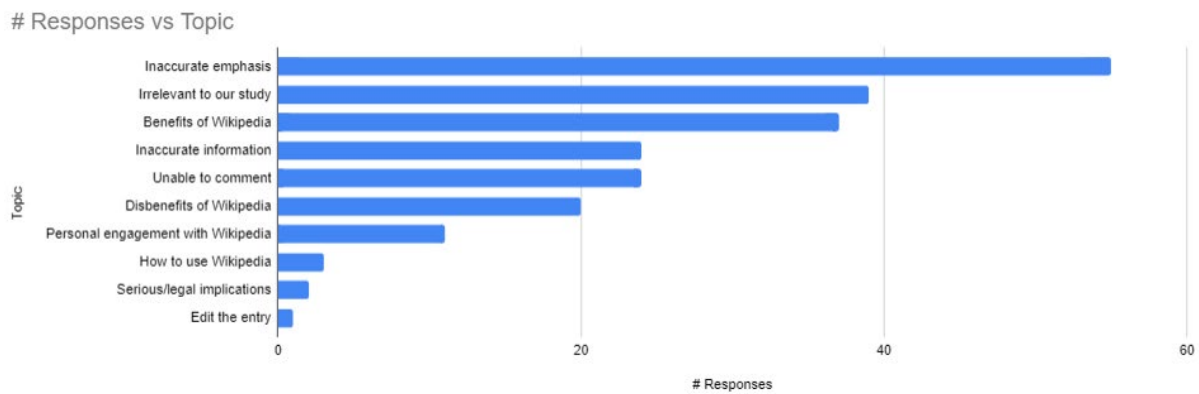

Figure 3: Free-text results analysis

# Discussion

The primary aim of this study was to collect and analyse researcher sentiment towards the Wikipedia citation of their article or book. Our results suggest there is general trust among researchers in Wikipedia both in terms of representativeness and accuracy. Most would also recommend the Wikipedia page where their work is cited to a colleague or the general public.

The survey questions were chosen to solicit as many responses as possible: ensuring that the survey was quick to complete was a major consideration. Expectations of the survey were that it would capture a range of views indicating how researchers engage with Wikipedia and situated the need for this study against the backdrop of a growing body of research about Wikipedia's uses and abuses, the growth in Wikipedia citations and the use of Wikipedia as a teaching and research tool [14]. The response rate to our survey was 3.43%, with most responses engaging fully with both multiple choice and free-text survey questions within two weeks of the survey being launched. While the response rate alone was encouraging, the number of respondents who were willing to continue the conversation beyond the survey was notable. In response to Question 6: *Would you be willing to be contacted again and to be part of a bigger project on Wikipedia trust and public engagement*? Over 50% of respondents answered 'Yes' and provided contact details. As such the survey exceeded our expectations in gathering a secondary list of scholars with whom to explore trust in Wikipedia citations to their work. Although the survey design was designed to encourage completion, by being deliberately short (seven questions), the extent to which published authors would complete the survey was not anticipated.

As the main results suggest, researchers' views about the accuracy of the citations to their work, whether they would recommend the Wikipedia article about their work and if the article represented the original research topic well, were all positively indicated. There were interesting differences though between articles and books, with consistently higher scores for the article group throughout the survey (Table 4); with most being statistically significant.  Major studies indicate that researchers rate journal articles more highly as a source of trusted information [15].In asking about the accuracy of the citation in Wikipedia, data from the survey does not reveal anything about the reliability or quality of the information in a book vs article.

A potential reason for lower scores in the book group is that book citations in particular are often to works that are only marginally about the subject of the Wikipedia article; even though they may include a term related to the article in the title. In other words, there is no evidence that the author of the Wikipedia article has read the whole book or if they have just skimmed selectively. This may be a product of reading behaviour, or it may reflect most of the citations being added by "professional" Wikipedia editors (i.e. who are being given access via The Wikipedia Library) being generalists rather than specialists and therefore inserting citations based on fairly uncritical full-text searching. Exploring how much Wikipedia Editors engage in deep reading to construct the knowledge in the database or if they are primarily engaged in a "janitor" role [16] , making sure some type of citation gets added, was beyond the scope of this study. However, the questions arising from the respondents' comments about "inaccurate emphasis" suggest that there is a high level of clean-up of Wikipedia records taking place, rather than a deep textual analysis. The guidance provided to Wikipedia editors [17] urging full engagement with source material before discussing that content in an article also suggests that the practices of Wikipedia editors are less engaged with the detail of source material than the citation to the source itself:

> *It's a small Wikipedia page and it's likely mostly correct but doesn't offer a lot of information. I'd have to dig deeper to know if it's all accurate or thorough. I do appreciate it citing one of the charts in my book, however! (Book Author)*

Some respondents also noted that OA publication could be a contributing factor to increasing public engagement with published research. Overall, data from our survey did not substantiate claims that OA publication leads to greater citation via Wikipedia (as referenced in the in a study [11] and because most articles cited were behind a paywall, see Table 4). The free-text results do reveal some researcher sentiment connecting OA publication to greater citation potential and that this would be a benefit to researchers, potentially enhancing their trust of Wikipedia:

> *It would be nice [if] Wikipedia cited articles were open to the public. (Journal author)*

There are also some interesting differences when comparing discipline scores. Results suggest that among the six disciplines included in the publications covered by the survey, Physics & Mathematical Sciences (P&MS) are the most likely to refer a Wikipedia page to both a colleague and the public, more broadly, with Social Sciences and Humanities being the least likely to do the same. Together with Engineering and Technology, the P&MS group had also higher scores for perceived *accuracy* and representativeness of their research. The survey responses don't provide a clear indication as to why perceived accuracy in P&MS and Engineering and Technology is higher than other disciplines and there is little direct research evidence that indicates why perceived *accuracy of Wikipedia citations would* be higher in some disciplines over others. However, as research in P&MS and Engineering and Technology disciplines relies primarily on more quantitative and data-driven methodological approaches than research in the humanities, for example, the perceived *accuracy* may be attributable to researchers in those disciplines feeling more confident in the accuracy of a citation to their work that has been arrived at via more quantitative methods [18,19].

Analysing respondents' free-text entries in relation to the benefits perceived from Wikipedia citation, specifically where the respondent had scored 7 and above in relation to Q2 ("*The citation accurately represents the content of my publication*"), it becomes clearer that Wikipedia citation provides multiple benefits beyond citation *accuracy*. In the following sample quotes, the concept of trust is invoked in distinct ways:

> *Wikipedia seems to now be patrolled by trusted watchers, and it is possible to set alerts when content that may cite you or refer to you is altered - good way to check it yourself. (Journal author)*

> *Yes, I use Wikipedia in my research. I don't quote the articles themselves, but often use them to gain a broader understanding of a certain topic knowing that there are some caveats. The most important part of a Wikipedia page for me often is the references, which lead me to original source material. I'd rather use that than anything on Wikipedia. (Book author)*

As the option to add free text in response to Questions 2 and 7 was unrestricted by word count we also received a number of lengthy replies asking for edits to the Wikipedia entry and setting out in detail how the article should be re-written. The responses indicate that *accuracy* and *purpose* of citation in the Wikipedia entry are the main concerns of the respondents. Wikipedia citations to published works are topic driven and don't interrogate the topic in great depth but do add the validated source citation via DOI. The classification of the free-text comments indicates 77 "inaccurate *emphasis*" over 28 "inaccurate *information*"; respondents' free-text comments also back up the point that context is missing from the Wikipedia articles and that the purpose of Wikipedia is not as a primary source of information, but a "funnel" into that content.

> *In the context of research, Wikipedia is a funnel, not a source. That is, it helps finding in a single place an aggregate of references, with a commentary to facilitate their digestion. Ultimately, it's the peer-reviewed publications it cites that are the best guarantees of knowledge, and it's the responsibility of the researcher to adequately judge and evaluate each source they use. Consequently, it's important to understand the "trust" in Wikipedia as a gateway, more than as a final arbiter. (Journal Author)*

Even when researchers/authors disagreed with the context (*inaccurate emphasis*) in which their work had been cited, adding research citations to Wikipedia, was thought to provide a beneficial entry point to validated research,:

> *In research I will use it mainly to follow leads concerning individuals who come up elsewhere in my research - i.e. to see if they were prominent in some way. (Journal author)*

The benefits of the Wikipedia "funnel effect" for researchers were also not overlooked by researchers, with one respondent noting specifically how citations to their work in Wikipedia are developing relationships with the media to support more public engagement with their work:

> *Thank you for alerting me to the link, citing my article. I am calling it to the attention of a journalist contact who is preparing to write on a related subject. (Journal author)*

Finally, some authors also commented that citation in Wikipedia could potentially extend access and increase visibility of their work to a broader public. While limited in the free-text responses, one respondent linked citation in Wikipedia to the potential for OA publication to increase public access to research:

> *The Wiki article is a stub and could be lengthened considerably, but at least the page exists. Fiji soccer has a small online footprint. The citation shows the power of open access for topics of general interest. (Journal author)*

In summary, the free-text survey responses indicate a wide range of researcher sentiment about citation to their work in Wikipedia that can be broadly characterised as follows:

- Wikipedia seems to be a funnel into primary sources and rarely used to identify deep meaning about a topic (i.e. in comparison to a full, critical reading of the source material)
- Research citations added to Wikipedia may speed access to primary findings. This 'funnel effect' may be enhanced by the joint activities of researcher (i.e. through research and teaching activities) and TWL editor interventions (i.e. by adding and removing citations and writing articles based on validated research).
- Wikipedia research citations often distil global research, potentially overcoming regional biases and making access to and discussion of a broader range of global published research more equitable.
- Uses of Wikipedia in academia are increasing; the use of Wikipedia as a research and teaching tool supports the perception of Wikipedia as a trusted site of information
- Improving Wikipedia article and citation accuracy (i.e. researcher advocacy for Wikipedia) may reinforce the public message that they can trust the research being cited in Wikipedia
- Wikipedia citations may support public engagement with validated research and co-create researcher value in an increasingly open landscape

## Limitations

The research disciplines are categorised according to the manuscript Fields of Research classification and not as indicated by respondents. We acknowledge that, in some rare cases, there might be some respondents who are wrongly labelled in particular disciplines, however we strongly believe that our results rightly represent each respective field as most researchers publish within their own discipline.

The free-text responses, while a reasonably significant proportion of respondents and indicating the value of Wikipedia citations to researchers, have only been analysed at surface level. Further contextualisation of the benefits and disbenefits of Wikipedia citations is required and will be picked up in a further, more in-depth study with a sub-set of respondents.

## Recommendations

We acknowledge that this survey is only the first step in attempting to understand researchers' trust in Wikipedia citations to their work and would like to offer four recommendations for further consideration:

1) as the survey responses indicate, researchers perceived a high level of trust in the citation their work in Wikipedia. Investing in further studies about how the inclusion of more accurate research citations in Wikipedia could enhance *public trust* in research could further improve researchers' perception of the benefits of a citation to their work in Wikipedia.

2) as researchers are primarily engaging with Wikipedia as a *teaching and research* tool, there may be opportunities to encourage specialist researcher-authors to update Wikipedia articles where they have expertise and as such improve the accuracy of citations in Wikipedia. However, we recognise that the labour associated with monitoring the accuracy of and editing Wikipedia citations is not insignificant for already overburdened researchers.

3) as the study was conducted by a team of publishing and publishing-adjacent bibliometric organisations, exploring what shifts in the research and publishing landscape need to occur could further enhance the perception of Wikipedia as a trusted source of research information. Publishers and their bibliometrics, data-analytics partners can support a *shift in mindset.*

4) while publishers can make efforts to open their catalogues to vetted Wikipedia editors to increase accuracy of citations, some further investment in tools and system solutions that alleviate the burden on time-poor researchers is required.

Opening up catalogues to a broader range of published research is one measure that publishers can adopt to enhance researcher perceptions of trust in Wikipedia citations. More significantly, perhaps, is establishing the benefit of Wikipedia citations in the research evaluation process. Giving special consideration to where humans are involved in the creation, curation and distribution of knowledge and noting how "alignment" between humans and machines in the context of Large Language Models (LLMs) and the future of Wikipedia could continue to build trust [20]:

> *And if you appreciate the editors of Wikipedia are human, they have human motivations and concerns and that their motivations are providing high-quality educational material to align with your needs, then you can essentially put trust in the system.*

## Conclusion

The survey responses from researchers indicate that there is a reasonably high level of interest in the Wikipedia citations to their published work. There also seems to be a general level of trust about the use of authors' published work in Wikipedia.

Where *"inaccurate emphasis"* had been introduced by the Wikipedia editors, researcher responses still indicated a level of trust in Wikipedia citations, primarily the citation serves as a funnel into the source material. This finding underscores the need for Wikipedia editors to continue adding research citations to articles and for initiatives like TWL to continue providing streamlined access to publishers' source content. While the survey results indicate that accurate citations added to Wikipedia contribute to its trustworthiness as a funnel into research, a more intense interrogation of the perceived researcher benefits of engaging with Wikipedia as a vital funnel into validated research is required, especially if Wikipedia is to be perceived more widely as a trusted source of information for research and teaching.

Finally, the level of author engagement with both the survey and Wikipedia suggests that publisher collaboration with The Wikipedia Library may be beneficial to researchers, increase trust in Wikipedia more generally and provide additional benefits by opening up research to a broader public audience:

> *I think Wikipedia is a great site and am very glad to see it being taken seriously by publishers.  It's a great space for public dialogue and presentation of ideas. (Book author, quote abbreviated)*

## Data Availability Statement

Summary data and statistical analyses are available on Figshare, 10.6084/m9.figshare.26003593 Descriptive metadata data was obtained from Altmetric and Dimensions under a research license.

## References

1. Bould MD, Hladkowicz ES, Pigford A-AE, Ufholz L-A, Postonogova T, Shin E, et al. References that anyone can edit: review of Wikipedia citations in peer reviewed health science literature. BMJ : British Medical Journal. 2014;348: g1585. doi:10.1136/bmj.g1585
2. Azzam A, Bresler D, Leon A, Maggio L, Whitaker E, Heilman J, et al. Why Medical Schools Should Embrace Wikipedia: Final-Year Medical Student Contributions to Wikipedia Articles for Academic Credit at One School. Academic Medicine. 2017;92. Available: https://journals.lww.com/academicmedicine/fulltext/2017/02000/why_medical_schools_should_embrace_wikipedia_.22.aspx
3. Wikipedia contributors. Academic studies about Wikipedia. Wikipedia, The Free Encyclopedia. 30 Sep 2023. Available: https://en.wikipedia.org/w/index.php?title=Academic_studies_about_Wikipedia&oldid=1177928248
4. Proffitt M. Leveraging Wikipedia: Connecting Communities of Knowledge. Chicago: ALA Editions; 2018.

5. Meta contributors. Community Insights/Community Insights 2021 Report/Thriving Movement. 27 Sep 2021 [cited 4 Oct 2023]. Available: https://meta.wikimedia.org/w/index.php?title=Community_Insights/Community_Insights_2021_Report/Thriving_Movement&oldid=25668269
6. Meta contributors. GLAM Wiki 2023. 23 Oct 2023 [cited 4 Oct 2023]. Available: https://meta.wikimedia.org/w/index.php?title=GLAM_Wiki_2023&oldid=25653721
7. Mitchell M. Leveraging Wikipedia: connecting communities of knowledge. Technical Services Quarterly. 2019;36: 226–227. doi:10.1080/07317131.2019.1584990
8. Proffitt M. Welcome our resident Wikipedian! 22 May 2012 [cited 4 Oct 2023]. Available: https://hangingtogether.org/welcome-our-resident-Wikipedian/
9. Annual Reviews and Wikipedia. Seeking a Wikipedian-In-Residence. 25 Feb 2020 [cited 4 Oct 2023]. Available: https://annualreviewsnews.org/2020/02/25/seeking-a-wikipedian-in-residence/
10. Yang P, Shoaib A, West R, Colavizza G. Wikipedia and open access. ArXiv. 2023. doi:10.48550/arxiv.2305.13945
11. Nicholson JM, Uppala A, Sieber M, Grabitz P, Mordaunt M, Rife SC. Measuring the quality of scientific references in Wikipedia: an analysis of more than 115M citations to over 800 000 scientific articles. FEBS J. 2021;288: 4242–4248. doi:https://doi.org/10.1111/febs.15608
12. R Core Team. R: A Language and Environment for Statistical Computing. Vienna, Austria: R Foundation for Statistical Computing; 2021. Available: https://www.r-project.org/
13. Wickham H, Averick M, Bryan J, Chang W, McGowan L, François R, et al. Welcome to the tidyverse. J Open Source Softw. 2019;4: 1686. doi:10.21105/joss.01686
14. McDowell ZJ, Vetter MA. Wikipedia as Open Educational Practice: Experiential Learning, Critical Information Literacy, and Social Justice. Soc Media Soc. 2022;8: 20563051221078224. doi:10.1177/20563051221078224
15. Nicholas D, Watkinson A, Volentine R, Allard S, Levine K, Tenopir C, et al. Trust and Authority in Scholarly Communications in the Light of the Digital Transition: setting the scene for a major study. Learned Publishing. 2014;27: 121–134. doi:10.1087/20140206
16. Sundin O. Janitors of knowledge: constructing knowledge in the everyday life of Wikipedia editors. Journal of Documentation. 2011;67: 840–862. doi:10.1108/00220411111164709
17. Wikipedia Contributors. Wikipedia:Read the source. 21 Nov 2021 [cited 4 Oct 2023]. Available: https://en.wikipedia.org/wiki/Wikipedia:Read_the_source
18. Lakshman M, Sinha L, Biswas M, Charles M, Arora NK. Quantitative Vs qualitative research methods. The Indian Journal of Pediatrics. 2000;67: 369–377. doi:10.1007/BF02820690
19. Borrego M, Douglas EP, Amelink CT. Quantitative, Qualitative, and Mixed Research Methods in Engineering Education. Journal of Engineering Education. 2009;98: 53–66. doi:https://doi.org/10.1002/j.2168-9830.2009.tb01005.x
20. Jon Gertner. Wikipedia's Moment of Truth. 18 Jul 2023 [cited 4 Oct 2023]. Available: https://www.nytimes.com/2023/07/18/magazine/wikipedia-ai-chatgpt.html
21. Areia C, Burton K, Taylor M, Watkinson C. Research Citations Building Trust in Wikipedia. In progress. 2024. Data Supplement. figshare. Dataset. https://doi.org/10.6084/m9.figshare.26003593